\documentclass[sigconf]{acmart}
\pdfoutput=1 

\renewcommand\footnotetextcopyrightpermission[1]{} 
\usepackage{graphicx}
\usepackage{algorithm}
\usepackage{algpseudocode}
\usepackage{xcolor}
\usepackage{amsmath}
\usepackage{multirow}
\definecolor{orange}{rgb}{1.0,0.5,0.0}

\AtBeginDocument{%
  \providecommand\BibTeX{{%
    \normalfont B\kern-0.5em{\scshape i\kern-0.25em b}\kern-0.8em\TeX}}}

\setcopyright{acmcopyright}
\copyrightyear{2023}
\acmYear{2023}

\acmConference[KDD '23]{Make sure to enter the correct
  conference title from your rights confirmation emai}{June 03--05,
  2023}{Long Beach, CA}

\begin{document}

\title{Cooperating Graph Neural Networks with Deep Reinforcement Learning for Vaccine Prioritization}

\author{Lu Ling}
\email{ling58@purdue.edu}
\affiliation{%
  \institution{Purdue University}
  \streetaddress{305 N University St}
  \city{West Lafayette}
  \state{Indiana}
  \country{USA}
  \postcode{47906}
}

\author{Washim Uddin Mondal}
\affiliation{%
  \institution{Purdue University}
  \city{West Lafayette}
  \country{USA}}
\email{wmondal@purdue.edu}

\author{Satish V. Ukkusuri}
\authornotemark[1]
\affiliation{%
  \institution{Purdue University}
  \city{West Lafayette}
  \state{Indiana}
  \country{USA}
  \postcode{47906}}
\email{sukkusur@purdue.edu}







\begin{abstract}
\end{abstract}

\begin{abstract}
This study explores the vaccine prioritization strategy to reduce the overall burden of the pandemic when the supply is limited. 
Existing methods conduct macro-level or simplified micro-level vaccine distribution by assuming the homogeneous behavior within subgroup populations and lacking mobility dynamics integration. 
Directly applying these models for micro-level vaccine allocation leads to sub-optimal solutions due to the lack of behavioral-related details.
To address the issue, we first incorporate the mobility heterogeneity in disease dynamics modeling and mimic the disease evolution process using a Trans-vaccine-SEIR model. 
Then we develop a novel deep reinforcement learning to seek the optimal vaccine allocation strategy for the high-degree spatial-temporal disease evolution system. The graph neural network is used to effectively capture the structural properties of the mobility contact network and extract the dynamic disease features. 
In our evaluation, the proposed framework reduces 7\% - 10\% of infections and deaths than the baseline strategies.
Extensive evaluation shows that the proposed framework is robust to seek the optimal vaccine allocation with diverse mobility patterns in the micro-level disease evolution system.  
In particular, we find the optimal vaccine allocation strategy in the transit usage restriction scenario is significantly more effective than restricting cross-zone mobility for the top 10\% age-based and income-based zones.
These results provide valuable insights for areas with limited vaccines and low logistic efficacy.

\end{abstract}





\maketitle
\pagestyle{plain}

\section{Introduction}

The pandemic has posed unprecedented global health, economic and social challenges and consequently raised immediate concerns for effective vaccine allocation strategies~\cite{anderson2020will,nguyen2016optimizing,keeling2012optimal,sunohara2021effective}. However, the vaccine supply has been limited in various phases of the pandemic. For example, vaccines are in inadequate supply in the early outbreaks \cite{penney2021hot}; a limited amount of vaccine boosters are available between the initial outbreak and widespread control \cite{massonnaud2022evaluating}; vaccines are in short supply in developing countries that rely on vaccine donations from developed nations \cite{fidler2012negotiating}. Additionally, vaccine distribution faces challenges with logistics, trained personnel, and efficient scheduling. Based on these considerations, vaccines must be given to areas with the greatest need, thereby, prioritizing vaccines to curb the spread and severity of infections is of great importance to policymakers.




The study of vaccine prioritization strategy has drawn widespread attention\cite{chen2022effective,hao2022reinforcement,jain2021rapid}. Existing studies have explored the macroscopic vaccine prioritization strategies at the national, state, and city levels. Although the macroscopic strategies give an aggregated guide for the stockpiles and top-level vaccine planning, vaccine prioritization at the micro-geographical level would help monitor and adjust the long-term vaccine planning effectively. In particular, it is essential to influence its uptake for individuals via their nearby clinics. Indeed, studies \cite{sharma2020understanding,larson2014understanding} suggested that distance is a prime factor that influences the refusal or hesitancy of the vaccine for areas in South Africa and India. Besides, Mazar et al. \cite{mazar2022distance} found that being 0.25 miles vs. 5 miles from the vaccine site (CVS or Walgreens) was associated with a 9\% lower vaccination rate during COVID-19 in California. Those observations indicate that the behavior on a  micro-geographical scale is needed when designing the vaccine prioritization strategies. 


A major challenge in vaccine prioritization strategies is the simulation and prediction of disease propagation. In addition to the epidemiological factors, most studies focus on the impact of demographic characteristics such as age \cite{dalgicc2017deriving,chowell2009adaptive,sunohara2021effective} and occupations \cite{nunner2022prioritizing,islam2021evaluation,strodel2021covid,buckner2021dynamic} (e.g., essential workers recommended by the Centers for Disease Control and Prevention (CDC)). Some studies included additional features like comorbidity status \cite{tuite2010optimal}, pregnancy status \cite{lee2010computer}, and the contact pattern \cite{laskowski2015strategies,chowell2009adaptive,lee2010computer}, which is generally assumed to be homogeneous within the population subgroup. Meanwhile, the extensive social activities and mobility promote the spatial and temporal heterogeneity of disease propagation \cite{ling2021spatiotemporal,manzo2020halting,herrmann2020covid,qian2021connecting}. The activity contagions, such as work, school, entertainment, and engagement, and the travel contagions from close contact with other commuters via travel modes would significantly amplify the disease contagions risk and promote disease propagation. The interaction between mobility risk and disease dynamics complicates the spatiotemporal vaccine distribution. However, few studies have integrated the heterogeneity of mobility contact in disease dynamics when sleeking the optimal vaccine prioritization strategies. The reason might be that the micro-level mobility data is hard to obtain and couple with the disease dynamics leading to new challenges in understanding this problem.  

Current computational approaches in vaccine distribution can be segregated into three groups, including disease compartmental models \cite{bubar2021model,buckner2021dynamic,ndeffo2011resource}, agent-based models \cite{lee2010computer,laskowski2015strategies} (ABM), and deep learning models \cite{meirom2021controlling,wei2021deep}. 
These models assume homogeneity in population subgroups to decrease the computational burden \cite{lee2010computer,laskowski2015strategies,lee2021strategies,islam2021evaluation}. However, that comes at the cost of reduced prediction accuracy for disease propagation and the credibility to evaluate the vaccine allocation strategy. Although deep learning methods are able to model complex systems, they are sensitive to diverse disease parameters. More importantly, the interpretability and credibility of vaccine distribution from the deep learning methods may be greatly reduced without incorporating behavioral-related heterogeneity in disease dynamics modeling.

\textbf{Our contributions.} Given the challenges discussed and the need to reduce the pandemic burden when vaccine supply is limited, optimizing vaccine distribution is of significant interest to policymakers. Thus, studying micro-geographical level vaccine allocation strategies is crucial in determining where, when, and how many vaccines to distribute.

To address these issues, we first develop a Trans-vaccine-SEIR model (SEIR refers to the susceptible-exposed-infected-recovered model) by extending the prior Trans-SEIR model \cite{qian2021connecting} to incorporate the impact of vaccines. Our model describes how, in the presence of a certain vaccine distribution strategy in the census block level, the disease propagates in a temporally varying graph whose nodes represent the census zones in the city, and edges describe the mobility connections between them. We employ a Graph Neural Network (GNN)-based policy approximator to yield vaccine distribution strategy as a function of the current state of the graph and train the GNN via an actor-critic-based reinforcement learning (RL) algorithm.




There are two novelties in this study. First, we integrate micro-geographical level mobility into the disease dynamics modeling and propose the Trans-vaccine-SEIR model to mimic the disease propagation. Second, we propose a RL-GNN framework to find the optimal vaccine allocation strategy. Within the framework, the deep RL is used to find the optimal solution in a high-degree spatiotemporal disease evolution graph. In particular, the GNN is reagrded as a policy approximator that effectively capture the structural properties of the mobility contact network and extract the dynamic disease features. Our RL-GNN framework overcomes the limitations faced by SEIR and ABM optimization in addressing the complex system dynamic problem. 

 We verify the effectiveness of our disease simulator by the real-world data and show that the optimal vaccine allocation strategy from the RL-GNN framework reduces 7\% and 10\% of infections and deaths compared to the baseline strategies. We also examine the effectiveness of proposed RL-GNN framework in multiple non-pharmaceutical interventions (NPIs). The optimal vaccine allocation strategy outperforms the baseline strategies and decreases infections and deaths by 10\% and 17\%, respectively. These experiments consistently show our proposed framework's superiority and robustness under diverse mobility patterns. In particular, among various NPIs, we find the optimal vaccine allocation strategy in the transit system usage restriction  (defined as the bus, ride-sharing vehicle, taxi, and van) is more effective than restricting cross-zone mobility in the top 10\% oldest/youngest/highest income/lowest income zones. 



\section{Related Work}

Vaccine prioritization has sparked an unprecedented discussion during the pandemic, including disease dynamics modeling and the optimization of the vaccine allocation strategy.

It is well understood that disease dynamics modeling is the key to evaluating the vaccine prioritization strategy. Vaccine prioritization strategy can be classified into two groups to mimic the reality of disease propagation based on the study scope. Extensive studies addressed macroscopic level vaccine allocation strategies such as nation \cite{matrajt2010optimizing,sunohara2021effective,penney2021hot}, state \cite{strodel2021covid}, region \cite{tildesley2006optimal}, or city \cite{yuan2015optimal,nguyen2016optimizing} levels. The macroscopic vaccine allocation is essential for regional vaccine allocation planning. The federal-level vaccine allocation strategy in the US is centered around the framework developed by the National Academies of Sciences, Engineering, and Medicine (NASEM) 
\cite{national2020framework}. However, they strongly rely on homogeneity assumption in the subgroup of the population such as age \cite{russell2021should, han2021time,bubar2021model} and occupancy \cite{nunner2022prioritizing,islam2021evaluation,jain2021rapid}. They ignore behavior-related heterogeneity among individuals, which might introduce inaccuracy estimation. In addition to the macroscopic strategy, an increasing amount of micro-level vaccine allocation studies have drawn people's attention in recent years. The micro-level studies focus on the risk approximation using social network \cite{chen2022effective}, contact pattern \cite{lee2010computer,laskowski2015strategies,buckner2021dynamic,lee2021strategies}, and social distance \cite{dalgicc2017deriving,choi2021vaccination} for the population subgroup. Those methods are selected based on the trade-offs between prediction accuracy and computational complexity. To reduce the computational complexity, they either ignore the internal epidemiological dynamics or limit the spatial and temporal heterogeneity in physical mobility movement when modeling disease dynamics. 
However, non-epidemiological factors \cite{ling2021spatiotemporal,nguyen2022self}, such as mobility and activity, significantly influence disease propagation. Ignoring these factors would hurt the accuracy of disease propagation simulation and, thereby, the credibility of the vaccine prioritization strategy. 

Vaccine allocation optimization models have generated widespread attention. These methods can be separated into three classes. The first approach is based on deterministic or stochastic disease compartmental models \cite{han2021time,chapman2022risk,tildesley2006optimal}, which use a system of differential equations to represent the disease dynamics. This approach suits for small and simplified systems, and the optimization formulation can be solved by the closed-form solution \cite{tildesley2006optimal}, brute-force search method \cite{medlock2009optimizing}, and greedy strategies \cite{wilder2018preventing}. The second class approach is network or ABM models \cite{nunner2022prioritizing,lee2010computer,laskowski2015strategies,lee2021strategies}. These models allow more group-level behavior variations. Since optimizing the epidemic outcomes in these models is much harder than in the conventional SEIR model, they tend to simplify the network structure by assuming static network structure or homogeneity in the contact matrix when modeling the disease transmission rate. Besides, they usually construct limited strategies based on the subgroup population to facilitate computational complexity. The last class approach is based on the deep learning method such as deep RL \cite{wei2021deep,beigi2021application,hao2022reinforcement}. Deep learning models have their advantages in modeling complex systems. However, they are sensitive to parameters. Studies cooperate deep learning models with disease compartment models, such as SEIR, assume homogeneous behavior in population subgroups, and ignore the mobility dynamics in the disease compartment models. That might decrease the interpretation of the disease dynamics and the credibility of the vaccine allocation strategy. 

As suggested by studies \cite{nguyen2022self,mazar2022distance}, 
a micro-geographical level vaccine prioritization strategy is needed to improve the vaccination rate and monitor long-term vaccine planning effectively. However, few studies address the micro-level spatiotemporal vaccine allocation strategy. They ignore the impacts of mobility dynamics, which would significantly underestimate the disease propagation and introduce biases in strategy evaluation. Despite deep learning models increasing the search space and enhancing the computational efficiency in finding the optimal vaccine allocation strategy, they fall short in considering the heterogeneity of disease dynamics, hindering the interpretability of the results. In this study, we address the abovementioned issue in the proposed framework and examine the robustness of the framework with diverse mobility patterns in NPIs schemes.

\section{System Dynamics Framework}

This section presents the details of the problem formulation and the designed framework.


\subsection{Trans-vaccine-SEIR Formulation}
\subsubsection{Notation}
The notion and description for the Tran-vaccine-SEIR model can be seen in Table \ref{tab:Trans_vaccine_SEIR_para}.
\begin{table}[t]
    \centering
    \begin{tabular}{{p{0.05\textwidth}p{0.4\textwidth}}}
    \hline
         Notation& Description variables\\ \hline
         $S_{ij}$& Susceptible population who are residents of zone $i$ and currently in zone $j$. Similar notation $E_{ij}$: exposed (latent) population, $I_{ij}$ infected population, $R_{ij}$: recovered population, $D_{ij}$ death population. \\
         $S_{i}^t$& Susceptible population who are residents of zone $i$. $t \in (u,v)$, $u$ refers to vaccinated, $v$ refers to non-vaccinated. Similar notations are used for $E$, $I$, $R$, and $D$ population\\
         $I_{a,i}^t$, $I_{s,i}^t$ & Vaccinated ($t=v$) and non-vaccinated ($t=u$) infected population at zone $i$ those who are asymptomatic and those who present symptoms. \\
         & \textbf{Fixed parameters}\\
         $N_{i}$ &Total population of zone $i$ \\ 
         $d$ & Travel mode, including low and median capacity travel modes\\
         $\beta^a$ & Disease transmission rate per valid contact during activity\\
         $\beta_d^t$ & Disease transmission rate per valid contact during travel using mode $d$\\
         $c^d$ & The ratio of people who choose travel mode $d$\\
         $\frac{1}{\sigma}$ & The expected latent duration for people remaining in $E$ before moving to $I$\\
         $\frac{1}{\gamma_e}$  & The expected recover duration for infected individuals who present symptoms($e \in s$) and who are asymptomatic ($e \in a$) remaining in $I$ before moving to $R$\\
         $k_{ij}$ & Expected number of valid contacts for residents of $i$ who are currently at zone $j$\\
         $k_{ij,kl}^d$ & Expected number of valid contacts for travelers from $i$ to $j$ who come across with travelers from $k$ to $l$ using the same travel mode $d$\\
         $g_i$ & Total departure rate of zone $i$\\
         $m_{ij}$& The rate of movement from zone $i$ to zone $j$, where $\sum_jm_{ij}=1$\\
         $r_{ij}$ & The return rate from zone $j$ to zone $i$\\
         $f_i$ & Activity contagions risk at zone $i$\\
         $f_{ij}$ & Activity contagions risk of residence of zone $i$ currently in zone $j$\\
         $h_{\overrightarrow{ij}}$,$h_{\overleftarrow{ij}}$ & Traveling contagions risk from zone $j$ to zone $i$ and from zone $j$ to zone $i$\\
         $\delta$ & Vaccine effectiveness\\
         $q$ & Clinical fraction in the percentage of infected population present symptoms\\
         $IFR$ & Infectious fatality rate\\
         $f_a^t$ & relative contagiousness of truly asymptomatic individuals for vaccinated ($t \in u$) and non-vaccinated ($t \in v$) people\\
         $V_i$ & Assigned vaccine in zone $i$ \\
         \hline
    \end{tabular}
    \caption{Variables and parameters in Trans-vaccine-SEIR formulation}
    \label{tab:Trans_vaccine_SEIR_para} \vspace{-10mm}
\end{table}


\subsubsection{Model details}

In this study, we address mobility dynamics' impacts on disease propagation in urban areas. In particular, we incorporate the mobility contact patterns between different census block zones into our model and circumvent the overly-simplified assumption of homogeneity used in the conventional SEIR model. We integrate micro-level mobility heterogeneity into the disease dynamics by following the prior Trans-SEIR model \cite{qian2021connecting}. Infected individuals spread the virus through physical contact during travel to the destination and activities they perform after reaching there. Travel and activity amplify the probability of exposure to the disease. Thus, we denote as travel and activity contagion risks respectively. We define two contagion periods each day. The first contagion period occurs when the residents of zone $i$ are influenced by activity contagion within zone $i$ and the travel contagion for the residents of zone $i$ traveling to zone $j$. The second contagion period occurs the residence of zone $i$ is influenced by activity contagion within zone $j$ after they travel to zone $j$ and the travel contagion for them traveling back to zone $i$.

\begin{figure}[t]
    \centering
    \includegraphics[width=\linewidth]{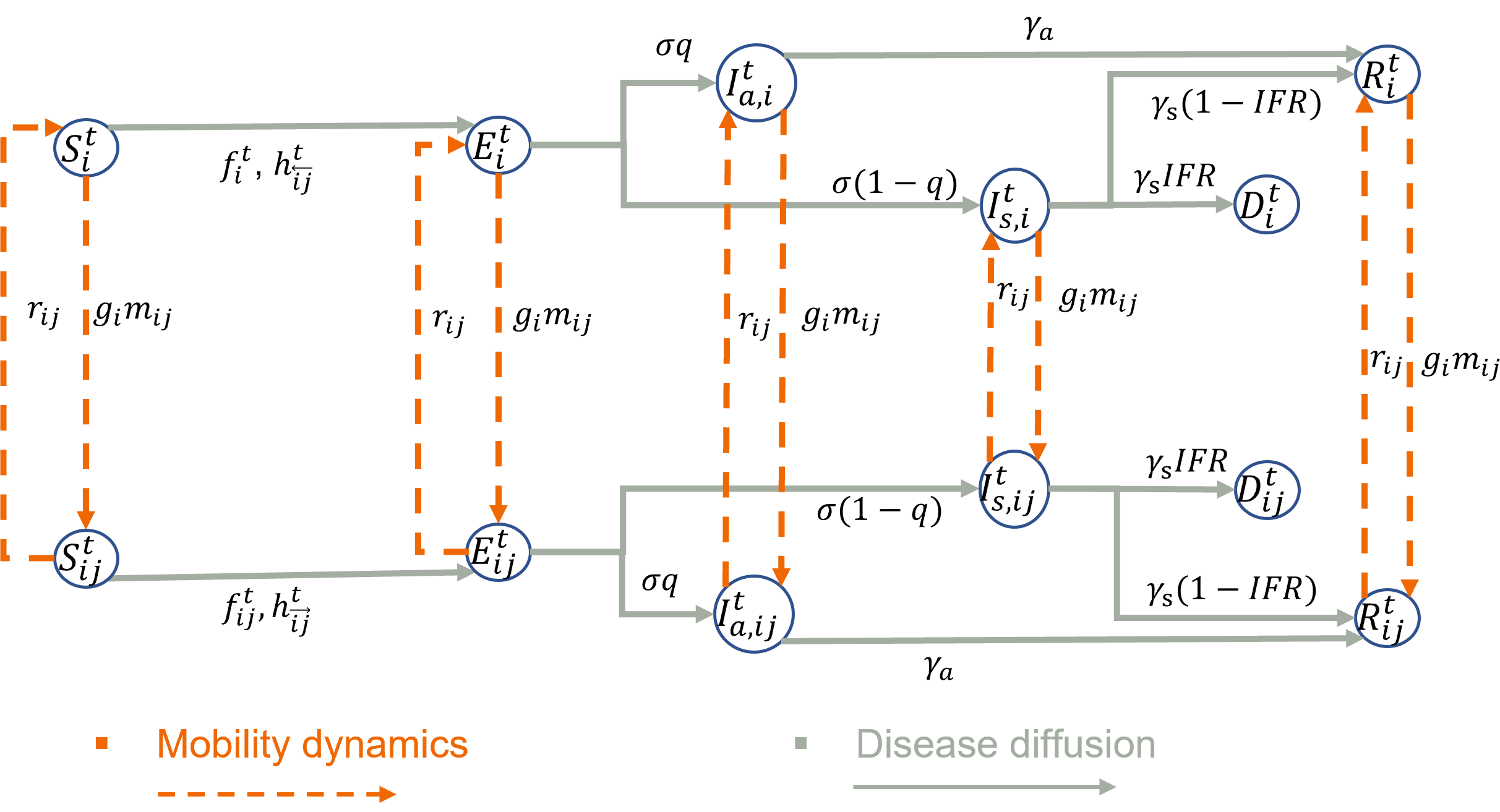}
    \caption{Trans-vaccine-SEIR model: mobility contact risk based disease dynamics under vaccine treatments}
    \label{fig:mobility_disease} \vspace{-4mm}
\end{figure}

We extend the Trans-SEIR model into the Trans-vaccine-SEIR model by incorporating the effect of the vaccine. The superscript $u$ represents the non-vaccinated states, and $v$ represents the vaccinated states. Susceptible individuals ($S$), comprising vaccinated $S^v$ and non-vaccinated $S^u$ individuals, are healthy individuals who have not been exposed to the disease. The susceptible individuals $S$ would shift to exposed ones $E$ during daily activity contagions risk $f$ and travel contagions risk $h$. The exposed individuals $E$, including vaccinated individuals $E^v$ and non-vaccinated individuals $E^u$,  are those exposed to the virus and not infectious. They would shift to the infectious state until the end of the incubation time $\frac{1}{\sigma}$. Then, infected individuals are either symptomatic ($I_s^u$ or $I_s^v$) or asymptomatic ($I_a^u$ or $I_a^v$). The parameter $q$ represents the ratio between symptomatic and asymptomatic infected individuals. Last, symptomatic individuals ($I_s^u, I_s^v$) are either recovered ($R^u, R^v$) or dead ($D^u, D^v$) after $\lambda_s$ time. The infected fatality ratio (IFR) represents the death ratio for infected individuals. Asymptomatic individuals ($I_a^u, I_a^v$) would recovery form the disease ($R^u, R^v$) with a recovery rate $\frac{1}{\lambda_a}$. In the Trans-vaccine-SEIR model, the mobility dynamics capture population movement across regions and disease parameters governing the population transition following the contagion process. The overview of the Trans-vaccine-SEIR framework is presented in Figure~\ref{fig:mobility_disease}. 

The mathematical formulation of the activity and travel contagion risks presents below.

\begin{equation}
\label{equ:activity1}
f_i=\beta^{a}k_i\frac{S_i[(f_{a,i}I_{a,i}+I_{s,i})+\sum_{k=1}^N(f_{a,k}I_{a,ki}+I_{s,ki})]}{N_i}
\end{equation}
\begin{equation}
\label{equ:activity2}
    f_{ij}=\beta^{a}k_{ij}\frac{S_{ij}[(f_{a,j}I_{a,j}+I_{s,j})+\sum_{k=1}^N(f_{a,k}I_{a,kj}+I_{s,kj})]}{N_j}
\end{equation}

\begin{equation}
\label{equ:travel1}
\begin{split}
    h_{\overrightarrow{ij}}=\sum_{d=1}^Dc_{ij}^d\beta_d^{t}(S_{ij})
    [\sum_{k=1}^p\sum_{l=1}^p\frac{k_{{lk,ij}}[(f_{a,i}I_{a,{lk}}+I_{s,{lk}})]}{N_{{lk}}}]
\end{split}
\end{equation}

\begin{equation}
\label{equ:travel2}
\begin{split}
    h_{\overleftarrow{ij}}=\sum_{d=1}^D c_{ij}^d\beta_d^{t}(S_{{ji}})
    [\sum_{k=1}^p\sum_{l=1}^p\frac{k_{{lk,ji}}[(f_{a,i}I_{a,{lk}}+I_{s,{lk}})]}{N_{{lk}}}]
\end{split}
\end{equation}

The equations \ref{equ:activity1} and \ref{equ:activity2} describe the activity contagions where health residents $S$ get infected by contacting infectious residents $I$ in zone $i$ and visitors from all other zones. The difference between equations \ref{equ:activity1} and \ref{equ:activity2} are the infected locations for healthy residents and the corresponding contact. The equation \ref{equ:travel1} and \ref{equ:travel2} give a mathematical representation of travel contagion. Equation \ref{equ:travel1} demonstrates the health population $S$ who get infected during travel while leaving their resident location for activities. Equation \ref{equ:travel2} represents health residents $S$ of zone $i$ get infected by infectious population $I$ in other places during travel. 

Then, the mathematical formulations of disease compartments in the Trans-vaccine-SEIR model are expressed below ($t \in (u,v)$). The susceptible population compartment presents in equation \ref{eqn:5}:
\begin{equation}\label{eqn:5}
\begin{split}
    \frac{dS_i^t}{dt}=\sum_{j=1}^Pr_{ij}S_{ij}^t-g_iS_i^t-f_i^t-\sum_{j=1}^Ph_{\overleftarrow{ij}}^t\\
    \frac{dS_{ij}^t}{dt}=g_im_{ij}S_i^t-r_{ij}S_{ij}^t-h_{\overrightarrow{ij}}^t-f_{ij}^t\\
\end{split}
\end{equation}

The exposure population compartment presents in equation \ref{eqn:6}:
\begin{equation}\label{eqn:6}
\begin{split}
    \frac{dE_i^t}{dt}=\sum_{j=1}^Pr_{ij}E_{ij}^t+\sum_{j=1}^Ph_{\overleftarrow{ij}}+f_i^t-g_iE_i^t-\sigma_iE_i^t\\
    \frac{dE_{ij}^t}{dt}=g_im_{ij}E_i^t+f_{ij}^t+h_{\overrightarrow{ij}}^t-r_{ij}E_{ij}^t-\sigma_iE_{ij}^t\\
\end{split}
\end{equation}

The infected population compartment presents in equation \ref{eqn:7}:
\begin{equation}\label{eqn:7}
\begin{split}
    \frac{dI_{a,i}^t}{dt}=\sum_{j=1}^Pr_{ij}I_{a,ij}^t+\sigma_i(1-q)E_i^t-g_iI_{a,i}^t-r_aI_{a,i}^t\\
    \frac{dI_{s,i}^t}{dt}=\sum_{j=1}^Pr_{ij}I_{s,ij}^t+\sigma_iqE_i^t-g_iI_{s,i}^t-r_sI_{s,i}^t\\ 
    \frac{dI_{a,ij}^t}{dt}=\sigma_i(1-q)E_{ij}^t+g_im_{ij}I_{a,i}^t-r_{ij}I_{a,ij}^t-r_aI_{a,ij}^t\\
    \frac{dI_{s,ij}^t}{dt}=\sigma_iqE_{ij}^t+g_im_{ij}I_{s,i}^t-r_{ij}I_{s,ij}^t-r_sI_{s,ij}^t\\
\end{split}
\end{equation}

The recovered population compartment presents in equation \ref{eqn:8}: 
\begin{equation}\label{eqn:8}
\begin{split}
    \frac{dR_i^t}{dt}=\sum_{j=1}^Pr_{ij}R_{ij}^t+r_aI_{a,i}^t+r_s(1-IFR)I_{s,i}^t-g_iR_i^t\\
    \frac{dR_{ij}^t}{dt}=g_im_{ij}R_i^t+r_aI_{a,ij}^t+r_s(1-IFR)I_{s,ij}^t-r_{ij}R_{ij}^t\\
\end{split}
\end{equation}

The death compartment presents in equation \ref{eqn:9}:
\begin{equation}\label{eqn:9}
\begin{split}
    \frac{d D_i^t}{dt} = \gamma_s IFR(I_{s,i}^t), 
    \frac{d D_{ij}^t}{dt} = \gamma_s IFR (I_{s,ij}^t)\\
\end{split}
\end{equation}

Equations \ref{eqn:5} - \ref{eqn:9} describe how mobility is involved in the disease dynamics as shown in Figure \ref{fig:mobility_disease}. 
These equations are consistent with the mobility contact from location $i$ $\longrightarrow$ $j$ $\longrightarrow$ $i$ and disease dynamics that follow the contagion process. 

\subsection{The RL-GNN Framework}
As described in the Trans-vaccine-SEIR model, we construct the census block zones in the city as a graph $G(t) =(V,E(t))$, where $V$ is a set of fixed nodes representing the zones in the city, $E(t) = \{e_{ij}(t)\}$ is the set of edges between $i$ and $j, (i, j)\in V$ at time $t$. The edge represents the mobility connection between zones. Each node $i$ is associated with disease compartment features $\zeta_i (t)$. 
They are random variables and vary in time. Each edge $e_{ij}(t)$ is associated with mobility-related disease compartment features  $\Psi_{ij}(t)$. 
The node and edge states at time $t$ depend on the state of itself, its neighbor, and the mobility interaction between its neighbors at time $t-1$. Please refer to the supplementary materials for node and edge feature details.


The objective is to find an optimal vaccine allocation strategy that minimizes the number of infected ($I, R, D$) and death population $D$ over time in a high-degree graph network. 
To seek optimality of the vaccine allocation in the high-degree graph with high dimensional node and edge features, we propose a RL-GNN framework and apply the deep RL approach to solve the system dynamics problem. RL\cite{sutton2018reinforcement} has the advantage of dealing with the uncertainty in a complex system and making decisions based on incomplete information as the environment changes, which suits for learning the optimal solution in the evolving graph.

\begin{figure}[t]
\centering
\includegraphics[width=\linewidth]{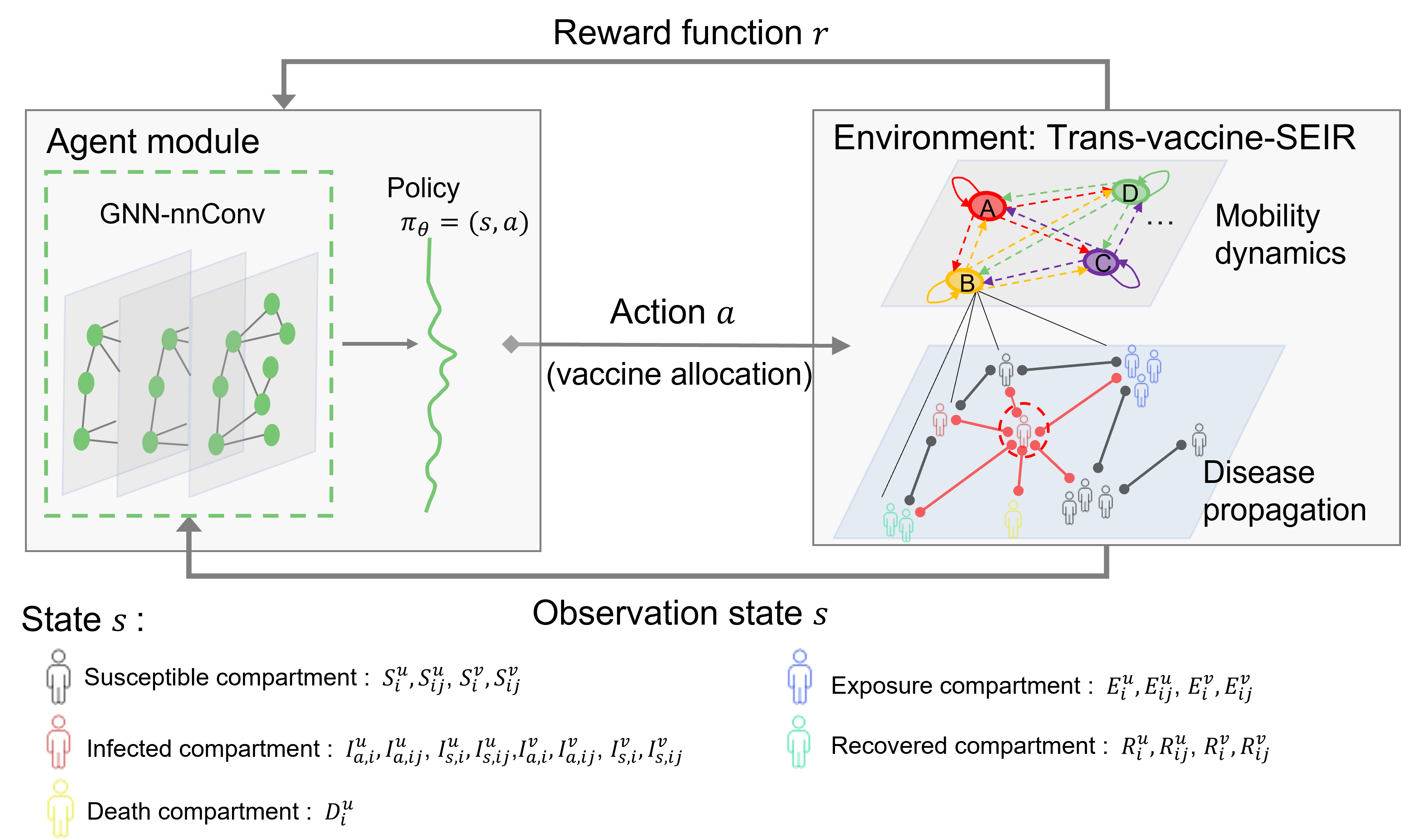}
\caption{The proposed controllable disease propagation framework}
\label{fig:framework}
\end{figure}

Figure~\ref{fig:framework} demonstrates the structure of the RL-GNN framework. Within the framework, the Trans-vaccine-SEIR module is the environment simulator, and the GNN is applied as the agent module. The GNN-based agent module first observes state $S(t)$ from the graph $G(t)$ environment in the Trans-vaccine-SEIR module. Based on the observation states, the GNN-based agent makes a vaccine allocation $\pi(t)$, which changes the graph environment's states. The Trans-vaccine-SEIR module mimics the disease propagation based on the mobility dynamics and vaccine impact, evaluates the vaccine allocation action from the GNN-based agent module, and outputs a reward $r(t)$. Then the proximal policy optimization (PPO) optimizer in the RL framework optimizes the action of the GNN-based agent based on the reward function $r(t)$. The architecture of the GNN-based agent module can be found in supplementary materials.

To improve the stability of the gradient descent direction in the PPO optimizer, we present our reward function:
\begin{equation}
\label{equ:reward}
\begin{split}
        R_t = \beta_E[(E_{t-1}-E_t)-(E^0_{t-1}-E^0_t)] + \\
        \beta_I[(I_{t-1}-I_t)-(I^0_{t-1}-I^0_t)] + \\
        \beta_R[(R_{t-1}-R_t)-(R^0_{t-1}-R^0_t)] +\\
        \beta_D[(D_{t-1}-D_t)-(D^0_{t-1}-D^0_t)]
\end{split}
\end{equation}
Where $E_t^0$, $I_t^0$, $R_t^0$, $D_t^0$ are the disease compartments for the baseline strategy, $E_t$, $I_t$, $R_t$, $D_t$ are the disease compartments based on the current vaccine allocation strategy decided by agent, $\beta_E$, $\beta_I$, $\beta_R$, $\beta_D$ are the coefficients for each disease compartments. 
In practice, $\beta_E = 1,\beta_I = 5, \beta_R = 100, \beta_D = 500$. 
Naively minimizing the daily increased number of infections ($I_t-I_{t-1}$, $R_t-R_{t-1}$, $D_t-D_{t-1}$) makes the optimization gradient noisy as the presence of infections for time $t$ has one-day or two-day delays in the simulator. Daily reduced exposed population ($E_t-E_{t-1}$) directly reflects the vaccine influence at time $t$ as the susceptible population $S_t$ can be protected by vaccine instead of transfer to the exposure period.
To stabilize the optimization gradient direction, we add the population-based vaccine allocation strategy as the baseline. In practice, this reward function design significantly improves the strategy's effectiveness.

Our setting differs from other work \cite{han2021time,chapman2022risk,wei2021deep,beigi2021application} in three critical aspects. First, we intergrate the census block level mobility dynamics in the conventional SEIR model to build the environment simulator and conduct vaccine allocation in the micro-geographic zones, which addresses the heterogeneity of mobility contact risk in disease evolution and provides a more realistic evaluation system for vaccine allocation. Second, we consider more detailed vaccine-related parameters (e.g., age-based vaccine effectiveness) and disease infection parameters (e.g., IFR, symptomatic ratio) based on demographic information. It captures the demographic-related population heterogeneity in disease transmission. Third, we do not assume a node can be quarantined from the graph. Isolating a high-degree node from the network might impair the transportation network quality and affect network connectivity.

\section{Trans-vaccine-SEIR model calibration}
We use the disease data in Marion county from January to August 2021 (during the COVID-19 period) as a case study. Zones at the census block level ensure the vaccine allocation strategy is in a high resolution for the general public. There are 253 census blocks in Marion county with a $N= 957,337$ population. The disease parameters are calibrated in Marion county before December 2020 to ensure non-vaccine available states. We classify the population into four age classes 0-14, 15-64, 65-74, and $75_+$ years old to capture the age heterogeneity of the disease parameters. The summary of parameter description and source can be found in supplementary materials.

\vspace{0.5cm}
\noindent\textbf{Infection.} As suggested by Thompson et al. \cite{thompson2021interim}, the vaccine effectiveness is not 100\%. The exposed individuals who are vaccinated $E^v$ might move into the infected compartment and start to spread the virus at the end of incubation time $\frac{1}{\sigma}$\cite{anderson2020reproduction} days. A clinical fraction $q$ of exposed individuals (including non-vaccinated and vaccinated yet to be protected) would show symptoms after being infected ($I_s^u, I_s^v$), while the others are asymptomatic ($I_a^u, I_a^v$). It's hard to obtain the asymptomatic rate due to multiple reasons (e.g., not reported). We follow the assumption \cite{islam2021evaluation} that fraction changes linearly with age such as $q_{i,a}=q_{i,75_+}-w(75_+-x_{i,a})$. The symptomatic rate for people over 75 years old in zone $i$ is $q_{i,75_+}\in (70\% ~ 100\%)$. In the simulator, we select the $q_{i,75_+}=85\%$. $w$ is the reduced rate for a one-year reduction in age, which we set into 0.7\cite{islam2021evaluation}. $x_{i,a}$ is the median age value in $a$ age group. Assuming $n_{i,a}$ is the number of individuals in the $a$ age group and $n_i$ is the total population in the census block $i$ . The symptomatic rate for the census block $i$ can be expressed as $q_i=\frac{\sum_aq_{i,a}n_{i,a}}{n_i}$. 
The mean of the symptomatic rate in Marion county is 59\%. The mean symptomatic rates for each age group are 0-14: 37\%, 15-64: 60\%, 65-74:81\%, and $75_+$: 85\%. The visualization of symptomatic rate in each census block zone can be found in the supplementary material.


We use the $f_a^u, f_a^v,f_s^u,$ and $f_s^v$ to represent the contagiousness of asymptomatic infected individuals who are non-vaccinated and vaccinated, the symptomatic infected individual who are non-vaccinated and vaccinated. 
We assume an average time for asymptomatic individuals and symptomatic individuals to recover as $\frac{1}{\gamma_a}$ and $\frac{1}{\gamma_s}$ days \cite{anderson2020reproduction}. Besides, we assume the virus will no longer infect individuals who have recovered from the disease within eight months. 

\vspace{0.5cm}
\noindent\textbf{Infection fatality rate.}
We use the age-based daily new cases divided by daily new deaths in Indiana to estimate the age-dependent infection fatality rate (IFR). The disease and demographic data are from April 2020 to July 2020 in Indiana. The statistic of age-based IFR in Marion county is $IFR_{0-14}=0, IFR_{15-64}= 0.009, IFR_{65-74}= 0.102$, and $IFR_{75_+}=0.263$. The census block $IFR_{i}$ will be adjusted based on the demographic information within the census block zone: $IFR_i =\frac{\sum_aIFR_{i,a}n_{i,a}}{n_i}$. $IFR_{i,a}$ is the IFR of the age group $a$ in state-level. 
The visualization of census block zone IFR in Marion county can be seen in supplementary materials.

\vspace{0.5cm}
\noindent\textbf{Transmission rate.}
The disease transmission rate is a function of reproduction number ($R_0$) and can be calculated by the next-generation matrices approach \cite{diekmann1990definition}. According to the Trans-SEIR model\cite{qian2021connecting}, the $R_0$ is upper-bounded by the the highest contagion rates for the travel segment and the activity location. We assume the activity transmission rate ($\beta^a$) and travel transmission rate ($\beta_d^t$) are homogeneous in Marion county and expressed as $\beta^a=\frac{R_0(\mu+\sigma)(\mu+\gamma)}{\sigma}, R_0\in (R_{min}, R_{max})$ and $\sum_d\beta_d^t=\frac{\beta^a}{2}, d\in(1,2)$. $d$ is the travel mode, which includes low-capacity travel mode (e.g., taxi and rider-sharing vehicle) and median-capacity travel mode (e.g., bus and van) in Marion county. The $R_0$ is estimated by COVID-19 cases from March 2020 to July 2021 in Marion county and can be found in supplementary materials.
Then, the transmission rate is estimated as $\beta^a\in(0.016-0.05), \beta_{d=1}^t\in(0.0006-0.002)$, and $\beta_{d=2}^t\in(0.008-0.025)$. 

\vspace{0.5cm}
\noindent\textbf{Vaccine allocation.}
The vaccine is assigned to non-vaccinated susceptible individuals $S^u$. That is $\frac{dS_i^u}{dt}=-v$, where $v$ is the amount of assigned vaccine. By assuming the vaccine effectiveness as $\delta$, the unprotected and protected susceptible population are $\frac{dS_i^v}{dt}=(1-\delta) * v$ and $\frac{dP_i^v}{dt}=\delta * v$.


\vspace{0.5cm}
\noindent\textbf{Mobility dynamics}
We adopt a large scale of mobile phone data in March 2020 to capture individual movement within Marion county. An effective trip is defined as a movement with distance larger than 20 meters. After trip extraction, we obtained 1812266 trips and 176856 users in total. The average repressiveness is 9.71\% in Marion county. The spatial variation of trip per person in the census block zone can be found in supplementary materials.


\section{Evaluation}
\subsection{Implementation Details}
\subsubsection{Environment simulator: Trans-vaccine-SEIR model}
\begin{figure}[t]
    \centering
    \includegraphics[width=\linewidth]{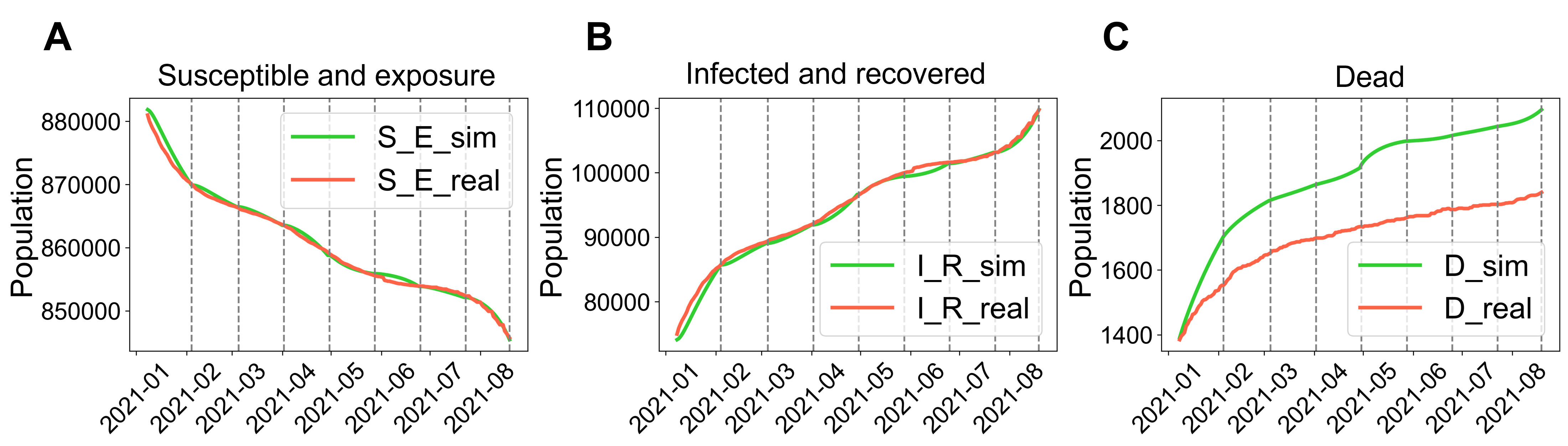}
    \caption{Environment simulator: Trans-vaccine-SEIR model; The suffix $\_sim$ and $\_real$ refer to the predicted and reality population in the disease compartments separately. $S\_E$ refers to the susceptible and exposure population. $I\_R$ refers to the infected and exposed population; $D$ refers to death population.}
    \label{fig:simulator_reality} \vspace{-3.5mm}
\end{figure}

To improve the interpretation of the disease dynamics, we propose the Trans-vaccine-SEIR model to demonstrate the disease propagation from both the epidemiological and non-epidemiological aspects. 
To mimic the disease propagation in Marion county from January to August 2021, we divide the study period to eight periods and simulate them continuously. The input data is the initial disease compartments in Marion county including $S, E, I, R, D$ and daily vaccine supply $V$. We apply the grid search to find the best parameters that fit the disease data in each period. Each period consists of four weeks. The first two weeks are used as the training dataset, and the last two weeks as the testing dataset. More details related with fitted parameters can be found in the supplementary materials. Besides, we construct three baseline approaches including population-based, even-based, and random-based. Population-based distribution means distributing vaccine in each census block zone according to the proportion of the census block population to the total population in Marion county. Even-based distribution means uniformly distribute vaccine to each census block zone. Random-based distribution means assigning vaccine to each census block zone follow random trials. The pseudo-code for the Trans-vaccine-SEIR model in the environment can be found in supplementary materials.
Figure~\ref{fig:simulator_reality} presents the predicted results and the realistic disease compartments. We define the accuracy metric as $C_{*} = \frac{*_{pred} - *_{real}}{*_{real}}$ for the ${*}$ disease compartment. The mean accuracy of the predicted disease propagation in each compartment is:  $C_{SE} = 0.0006$; $C_{IR} = 0.006$; $C_{D} = 0.002$. 




\subsubsection{RL and GNN Architecture}

We use Stable-Baseline3\cite{stable-baselines3} as our RL training framework.
We train our experiments using RTX 2080 Ti. The learning rate is 5.0e-5, the batch size is 8, the value function coefficient for the loss calculation is 0.5, the number of steps to run for each environment per update is 2, and the number of epochs when optimizing the surrogate loss is 10. Besides, we have two environment copies running in parallel. The GNN nnConv algorithm is implemented by pytorch-geometric \cite{Fey/Lenssen/2019}.

\subsection{Quantitative Evaluation}
In addition to three baseline vaccine allocation strategies, we simulate the non-vaccination scenario as a reference from January to August 2021. The well-trained optimal vaccine allocation strategy is used to compare baseline strategies. We use the improved case $IRD_{pro}$ and death $D_{pro}$ ratios as evaluation metrics. They are defined as follows:



\begin{equation}
\label{eqn:improved_case_ratio}
\begin{split}
        IRD_{pro} = \frac{(IRD^{no\_vac} - IRD^{opt})-(IRD^{no\_vac} - IRD^{*})}{(IRD^{no\_vac} - IRD^{*})} \times 100\% \\
        D_{pro} = \frac{(D^{no\_vac} - D^{opt})-(D^{no\_vac} - D^{*})}{(D^{no\_vac} - D^{*})} \times 100\% \\
\end{split}
\end{equation}

\begin{figure}[t]
    \centering
    \includegraphics[width=\linewidth]{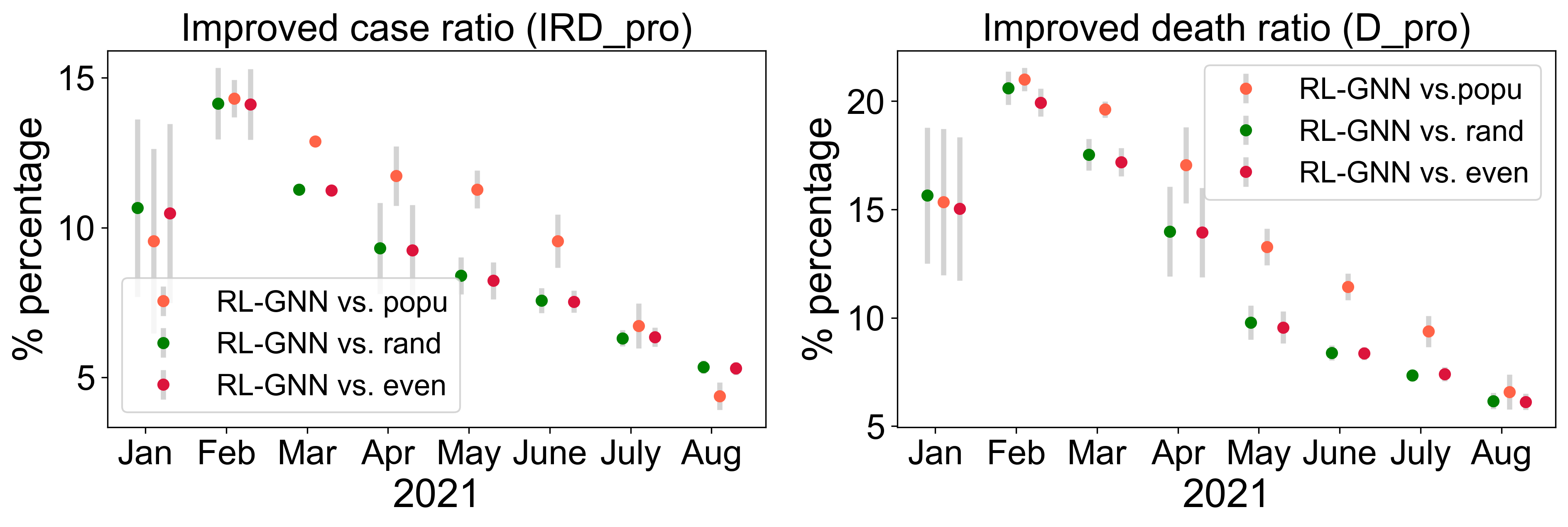}
    \caption{The time series evaluation of the optimal vaccine allocation strategy under RL-GNN framework}
    \label{fig:improved_ratio} \vspace{-4mm}
\end{figure}

Where the superscript $no\_vac$, $opt$, and ${*}$ represent the non-vaccination scenario, the optimal vaccine allocation strategy from the RL-GNN model, and the ${*}$ baseline strategy. $IRD$ and $D$ represent the total infected and death population. Figure~\ref{fig:improved_ratio} visualize the evaluation results in eight continuous periods. The optimal strategy's improved case and death ratios are 4\% -18\% and 6.7\%- 22\% than the baseline strategies. It indicates a significant improvement in reducing the number of infected and death populations for the proposed RL-GNN framework.

\subsection{Ablation Study}
We follow the standard practice and the same
protocol to assess the quality of the reward function in the RL framework and the GNN architecture in the agent module. 

\subsubsection{Reward function design}
We conduct several experiments to explore the reward function (See Eq.~\ref{equ:reward}). In addition, to evaluate each disease compartment, we consider three ways to explore the reward function: short-term baseline, long-term baseline, and no baseline. We use the population-based vaccine allocation strategy as the baseline to improve the gradient descent direction. The short-term baseline refers to a dynamic approach, adding the baseline in each time step $t$. Instead of adding the baseline each day, the long-term baseline is adding the baseline on the first day of the simulation. The improved case and death ratio of the proposed RL-GNN model using different metrics and baseline strategies in the reward function can be seen in Table \ref{tab:reward function}.

\begin{table}[ht]
    \centering
    \begin{tabular}{|c|c|c|c|}  \hline
       \multicolumn{2}{|c|}{Reward function} & \multirow{2}{*}{$IRD_{pro}$} & \multirow{2}{*}{$D_{pro}$} \\ 
       \cline{1-2}
        \multirow{1}{*}{with baseline strategy} & \multirow{1}{*}{metric} & & \\ \hline
         short-term &D& 4.04 & 24.31 \\
         short-term &I&17.45 & 23.09\\
         short-term &R&17.39 & 10.49\\
         short-term &E& 17.55&23.18 \\
         short-term &RD &15.32 & 19.17\\
         short-term &ED &10.02 & 21.13\\
         short-term &ID& 15.24&24.18 \\
        short-term &IRD& 17.15&24.04\\
         long-term & EIRD& -0.56&-1.06 \\
         no-baseline & EIRD &-0.54 & -1.81\\
         short-term & EIRD& 18.18&24.41 \\ \hline
    \end{tabular}
    \caption{Ablation study on reward function design}
    \label{tab:reward function}
\end{table}

\subsubsection{Agent architecture}
\begin{table}[t]
    \centering
    \begin{tabular}{|c|c|c|c|}  \hline
       \multicolumn{2}{|c|}{Agent architecture} & $IRD_{pro}$ & $D_{pro}$ \\  \hline
       \multicolumn{2}{|c|}{MLP} & 19.04 & 6.02\\
       \cline{1-2}
        \multirow{3}{*}{GNN} &GCNConv& 19.24&18.65 \\
         &CGConv& 4.99& 7.69\\
         &nnConv&19.78 & 23.02\\ \hline
    \end{tabular}
    \caption{Ablation study on agent architecture.}
    \label{tab:agentarchitecture} \vspace{-4mm}
\end{table} 
We apply two approaches to extract the features in the agent module: MLP and GNN. MLP embeds the features of the graph network in a fully connected way. Unlike the MLP, the GNN captures the properties of dynamic mobility contact network in the evolving graph. We compare three methods in GNN to embed the message passing for the nodes and edges' features: GCNConv\cite{kipf2016semi}, CGConv\cite{xie2018crystal}, and nnConv\cite{gilmer2017neural}. GCNConv only considers the node's feature embedding. CGConv uses the concatenation way to embed node features, neighboring node features, and edge features. nnConv applies a neural network to embed the node and edge features. The improved case and death ratios for different agent architectures are in Table \ref{tab:agentarchitecture}.

\section{Discussion}
The pandemic has resulted in unprecedented research worldwide to explore prevention policies\cite{brzezinski2020covid}, such as movement restriction, transit usage restriction, and vaccine prioritization strategies. Unfortunately, the complexity of inherent disease transmission and external factors make disease propagation hard to predict. In addition to the vaccine prioritization in a non-intervention scenario, evaluating the strategy associated with NPIs with diverse mobility patterns is important. To explore the capacity of RL-GNN framework for seeking the optimal vaccine allocation strategies by combining NPIs,  We propose four hypotheses related to demographic-based mobility restriction and transit usage restriction:

\begin{enumerate}
    \item The cross-zone mobility restriction is more effective in preventing disease propagation than the transit usage resection given the same vaccine allocation strategy.
    \item The optimal vaccine allocation for restricting cross-zone mobility in the top 10\% oldest zones would reduce more infected population than the optimal vaccine allocation for restricting cross-zone mobility in the top 10\% youngest zones.
    \item The optimal vaccine allocation for restricting cross-zone mobility in the top 10\% lowest income zones would reduce more death population than the optimal vaccine allocation for restricting cross-zone mobility in the top 10\% highest income zones.
    \item The effectiveness of the optimal vaccine allocation strategies is time invariant in diverse NPIs. 
\end{enumerate}

To test these hypotheses, we construct the mobility pattern in six scenarios based on the mobile phone data: (1) shifting the cross-zone travel for the top 10\% oldest and youngest zones to within-zone travel by assuming the same amount of trip generated by individuals and re-calibrate the travel and activity-related parameters; (2) shifting the cross-zone travel for the top 10\% highest and lowest income zones to within zone travel by assuming the same amount of trip generated by individuals and re-calibrate the travel and activity related parameters; (3) restricting the median capacity transit usage (defined as bus and vans), low (defined as taxi and rider-sharing vehicle) and median capacity transit usage. 

\vspace{0.5cm}
\noindent\textbf{The first hypothesis testing.} We follow the standard procedure to continuously simulate eight periods in the environment simulator. Instead of comparing the optimal vaccine allocation strategy in each NPI scenario, the comparison is solely based on the evaluation of diverse NPIs given the same vaccine allocation strategy. The evaluation metrics are defined as follows:
\begin{equation}
\label{eqn:simulation_eva}
\begin{split}
        IRD_{sim} = \frac{IRD^{pop}_{no\_rest}-IRD^{pop}_{*}}{IRD^{pop}_{no\_rest}},
        D_{sim} = \frac{D^{pop}_{no\_rest}-D^{pop}_{*}}{D^{pop}_{no\_rest}}
\end{split}
\end{equation}
Where $IRD_{sim}$ and $D_{sim}$ are the NPIs-based improved case and death ratio for the ${*}$ NPI. $IRD^{pop}_{non\_rest}$ and $IRD^{pop}_{*}$ are the infected population for population-based vaccine allocation strategy in the no mobility restriction scenario and the ${*}$ NPI scenario.

\begin{figure}[t]
    \centering
    \includegraphics[width=\linewidth]{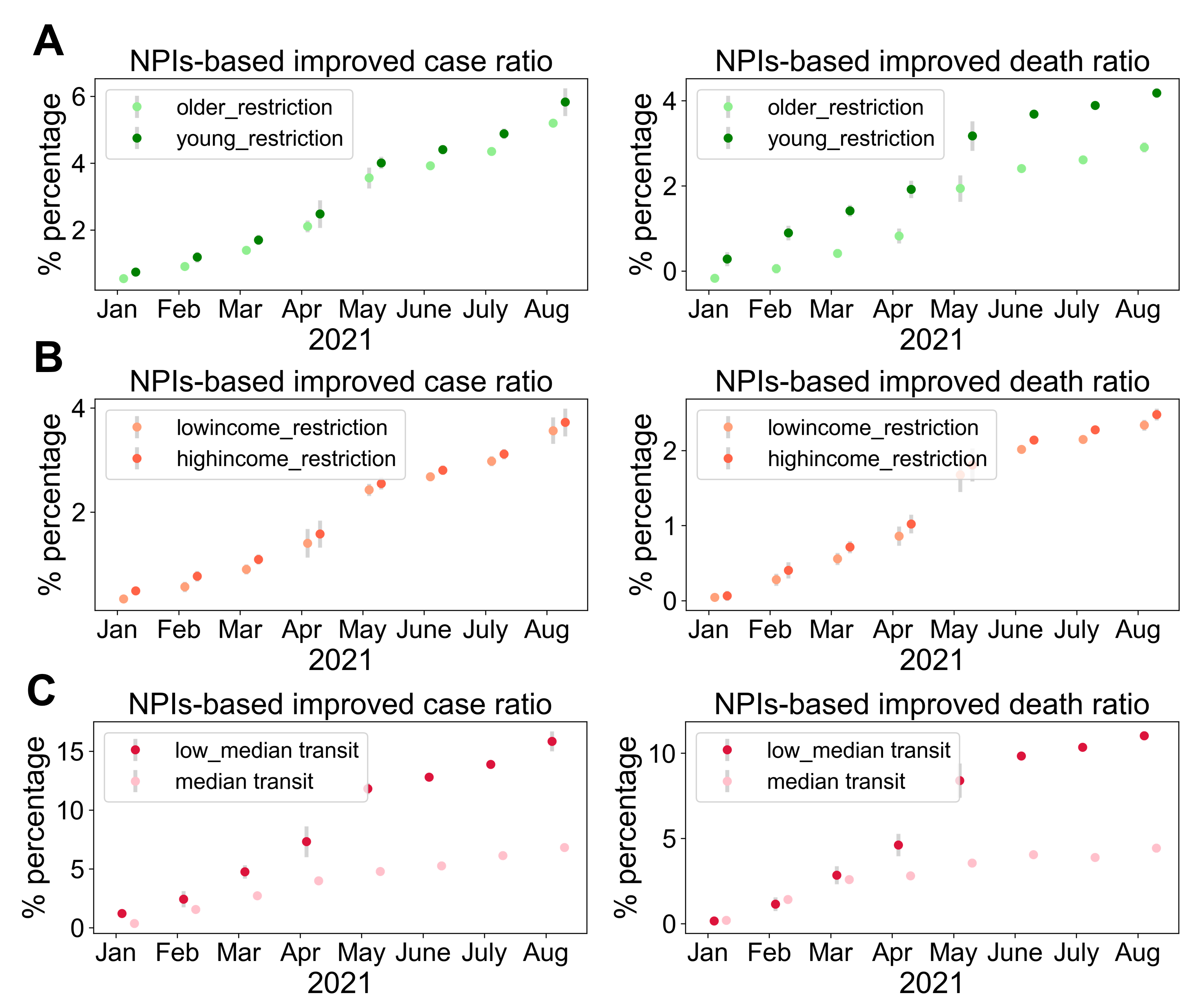}
    \caption{The time series evaluation of NPIs-based scenarios -- $IRD_{sim}$ and $D_{sim}$;A refers to the cross-zone mobility restrictions by age group. B refers to the cross-zone mobility restrictions by income groups. C refers to the transit usage restriction.}
    \label{fig:time_series_evaluation_simulator} \vspace{-3mm}
\end{figure}

\begin{figure}[ht]
    \centering
    \includegraphics[width=\linewidth]{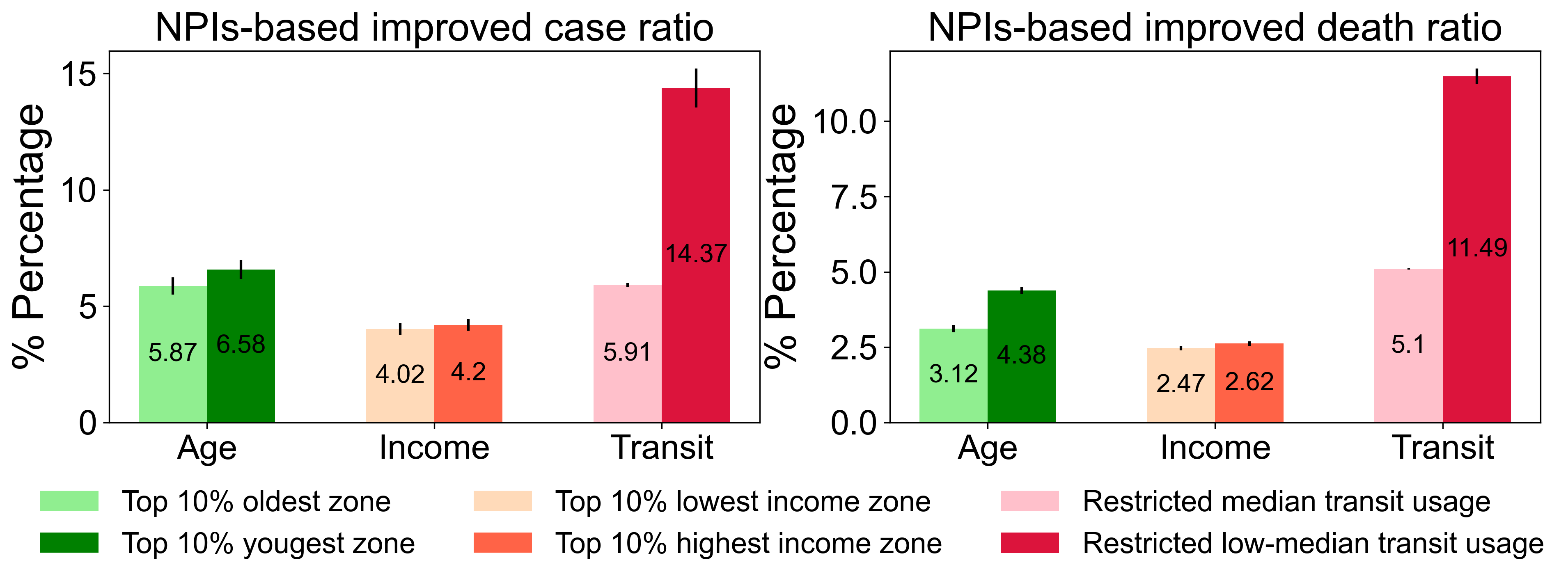}
    \caption{The evaluation of NPIs-based scenarios -- $IRD_{sim}$ and $D_{sim}$}
    \label{fig:simulator_sceniaor_comparsion}
\end{figure}

Figure~\ref{fig:time_series_evaluation_simulator} demonstrates the estimated results from January to August 2021 and shows that the transit usage restriction is more effective than the top 10\% age-based and income-based cross-zone mobility restrictions. Figure~\ref{fig:simulator_sceniaor_comparsion} demonstrates the NPIs-based improved infected case and death ratios at the end of the eight continuous periods. The NPIs-based improved infected case ratio for transit usage restriction is more than twice of other NPIs, which rejects the first hypothesis.

\vspace{0.5cm}
\noindent\textbf{The second, third, and last hypotheses testing.} We follow the standard procedure and train the vaccine allocation strategy based on the RL-GNN framework for diverse NPIs. Table \ref{tab:RL_GNN_simulation_compared_policies} shows the evaluation results of the optimal vaccine allocation strategies compared to three baseline strategies. The results show the robustness of the proposed RL-GNN model with diverse mobility patterns.

{
\begin{table*}[ht]
\renewcommand{\arraystretch}{1.2}
\setlength{\tabcolsep}{0.2em}
\footnotesize
\caption{The average improved case and death ratios in diverse mobility pattern}
\label{tab:RL_GNN_simulation_compared_policies}
\begin{tabular}{c|cc|cccc|cccc|cccc}
\hline
\multirow{3}{*}{Baseline} & \multicolumn{2}{c}{\multirow{2}{*}{No restriction}} & \multicolumn{4}{|c }{mobility restriction -   age}      & \multicolumn{4}{|c}{mobility restriction - income}              & \multicolumn{4}{|c}{transit usage restriction}                             \\  \cline{4-15}
                          & \multicolumn{2}{c}{}                                & \multicolumn{2}{|c}{older} & \multicolumn{2}{c}{young} & \multicolumn{2}{|c}{highincome} & \multicolumn{2}{c}{lowincome} & \multicolumn{2}{|c}{median capacity} & \multicolumn{2}{c}{low \& median capacity} \\ \cline{2-15}
                          & $IRD_{pro}$                     & $D_{pro}$                        & $IRD_{pro}$          & $D_{pro}$           & $IRD_{pro}$        & $D_{pro}$            & $IRD_{pro}$           & $D_{pro}$              &$IRD_{pro}$          & $D_{pro}$              & $IRD_{pro}$              & $D_{pro}$                 & $IRD_{pro}$                  & $D_{pro}$                   \\ \hline
population                       & 8.33+2.62                & 10.50$\pm$ 2.44               & 12.12+6.30   & 8.35+4.47  & 16.19+7.57  & 25.55+11.38 & 14.67+7.11     & 18.53+8.57    & 13.06+7.01    & 15.27+6.45    & 11.69+6.51       & 14.70+7.12       & 12.23+6.93           & 16.30+7.59          \\
even                      & 7.02+1.14                & 8.02+1.29                & 12.46+6.56   & 7.30+4.76  & 14.69+7.51  & 22.90+11.43 & 12.94+7.07     & 15.62+8.71    & 13.56+7.38    & 14.67+6.67    & 10.60+6.79       & 12.57+7.44       & 14.49+6.08           & 17.23+6.53          \\
random                    & 7.08+1.19                & 8.07+1.37                & 12.50+6.63   & 7.38+4.80  & 14.77+7.55  & 23.05+11.52 & 12.99+7.09     & 15.74+8.78    & 13.62+7.35    & 14.84+6.78    & 10.58+6.75       & 12.61+7.45       & 14.48+6.05           & 17.20+6.48      \\ \hline   
\end{tabular}
\end{table*}
}



To test the second and third hypotheses, we make a cross comparison for the age-based and income-based mobility restriction NPIs. Besides, we define the evaluation metrics as follow:

\begin{equation}
\label{eqn:optimal_eva}
\begin{split}
        IRD_{opt} = \frac{IRD^{opt}_{no\_rest}-IRD^{opt}_{*}}{IRD^{opt}_{no\_rest}},
        D_{opt} = \frac{D^{opt}_{no\_rest}-D^{opt}_{*}}{D^{opt}_{no\_rest}}
\end{split}
\end{equation}
Where $IRD_{opt}$ and $D_{opt}$ are the optimal NPIs-based improved case and death ratios from the optimal vaccine allocation strategy. $IRD^{opt}_{non\_rest}$ and $IRD^{opt}_{*}$ are the infected population for the optimal vaccine allocation strategy in the no mobility restriction scenario and the ${*}$ NPIs scenario.

\begin{figure}[t]
    \centering
    \includegraphics[width=\linewidth]{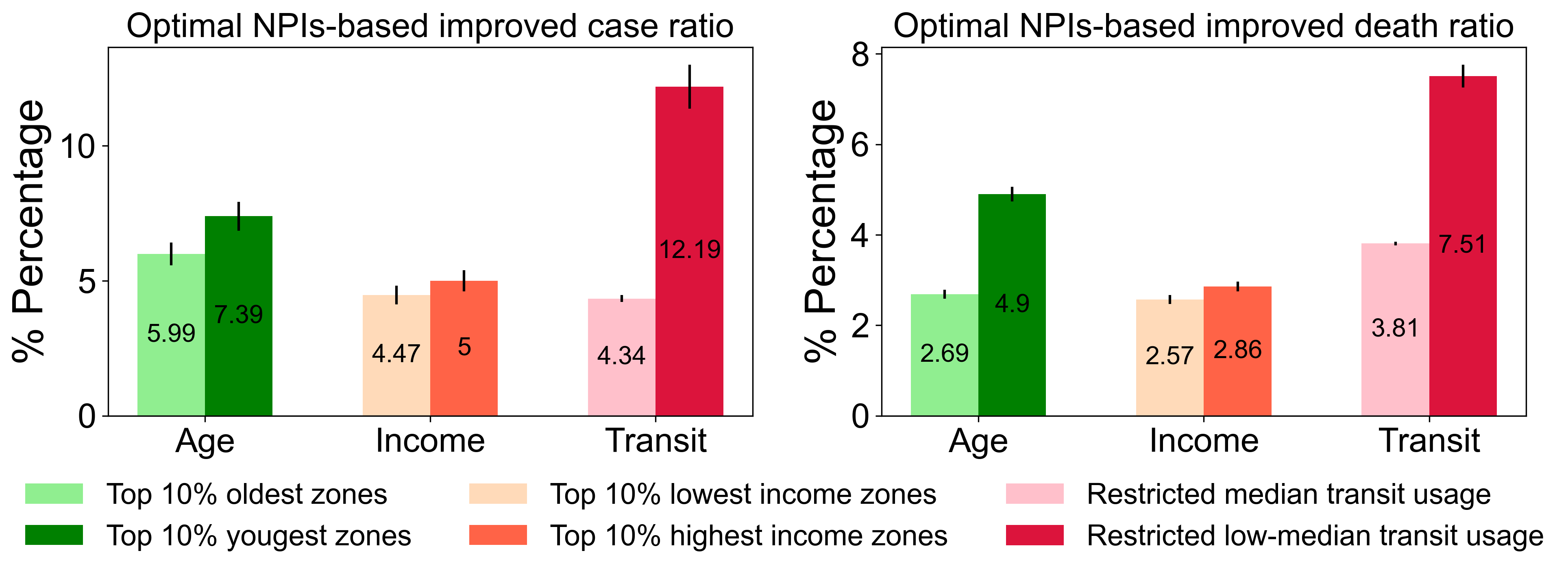}
    \caption{The evaluation of the optimal NPIs-based scenario -- $IRD_{opt}$ and $D_{opt}$}
    \label{fig:RLGNN_sceniaor_comparsion}
\end{figure}

Figure~\ref{fig:RLGNN_sceniaor_comparsion} visualizes the optimal NPIs-based improved infected case and death ratios at the end of the eight continuous periods. Based on the comparison, the optimal NPIs-based improved case ratios in restricting cross-zone mobility for the top 10\% oldest and youngest zones are 5.99\% and 7.39\%, separately. And the optimal NPIs-based improved death ratios in restricting cross-zone mobility for the top 10\% lowest income and highest income zones are 2.69\% and 4.90\%. That indicates restricting cross-zone mobility for the youngest zones and highest income zones is more effective than the oldest zones and lowest income zones, which rejects the second and third hypotheses.

\begin{figure}[t]
    \centering
    \includegraphics[width=\linewidth]{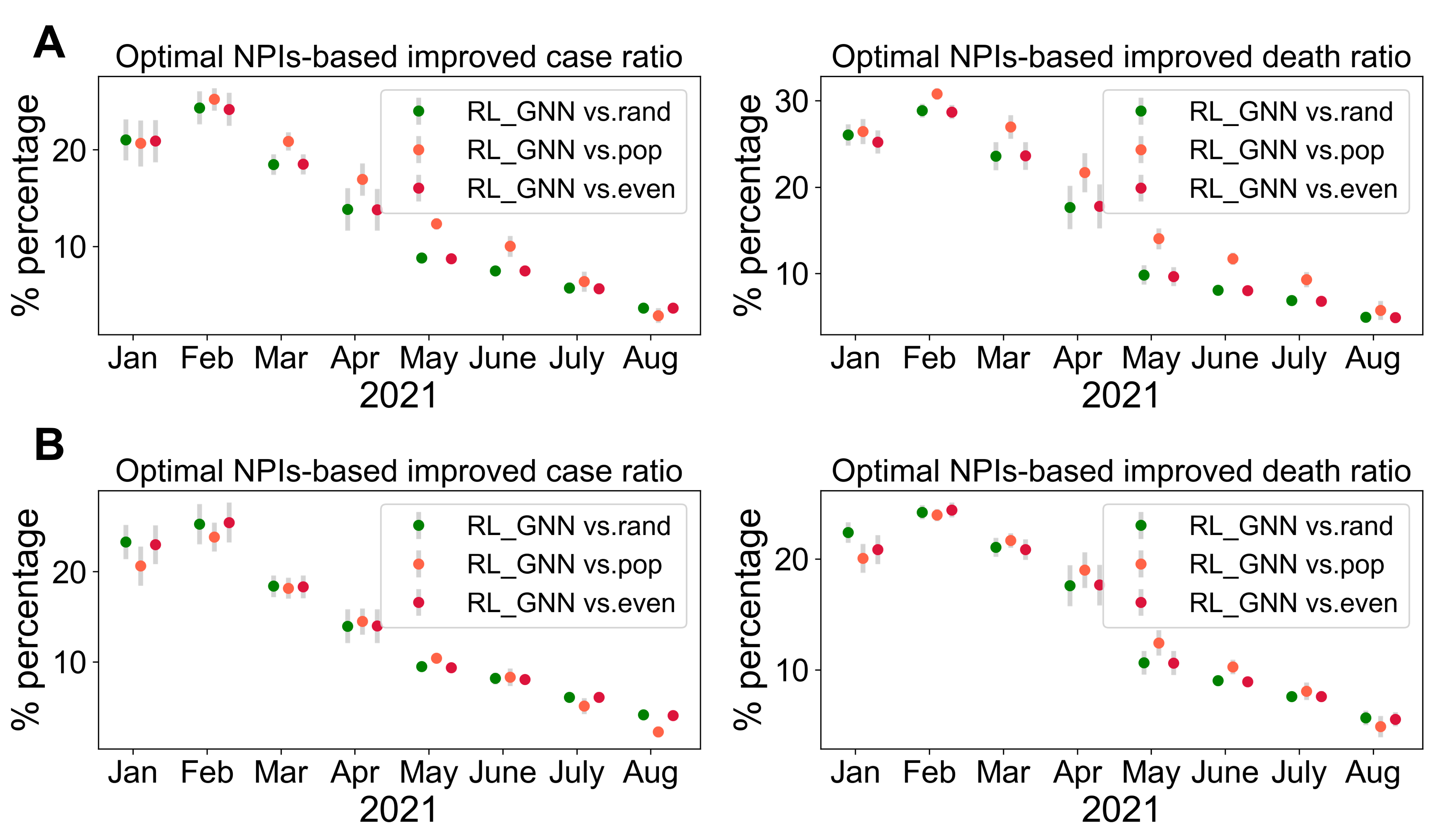}
    \caption{The time series evaluation of RL-GNN framework under the income-based mobility restriction --$IRD_{pro}$ and $D_{pro}$; A: restricting cross-zone mobility for the top 10\% highest income zones; B: restricting cross-zone mobility for the top 10\% lowest income zones.}
    \label{fig:RL_GNN_income_timeseries}
\end{figure}

\begin{figure}[t]
    \centering
    \includegraphics[width=\linewidth]{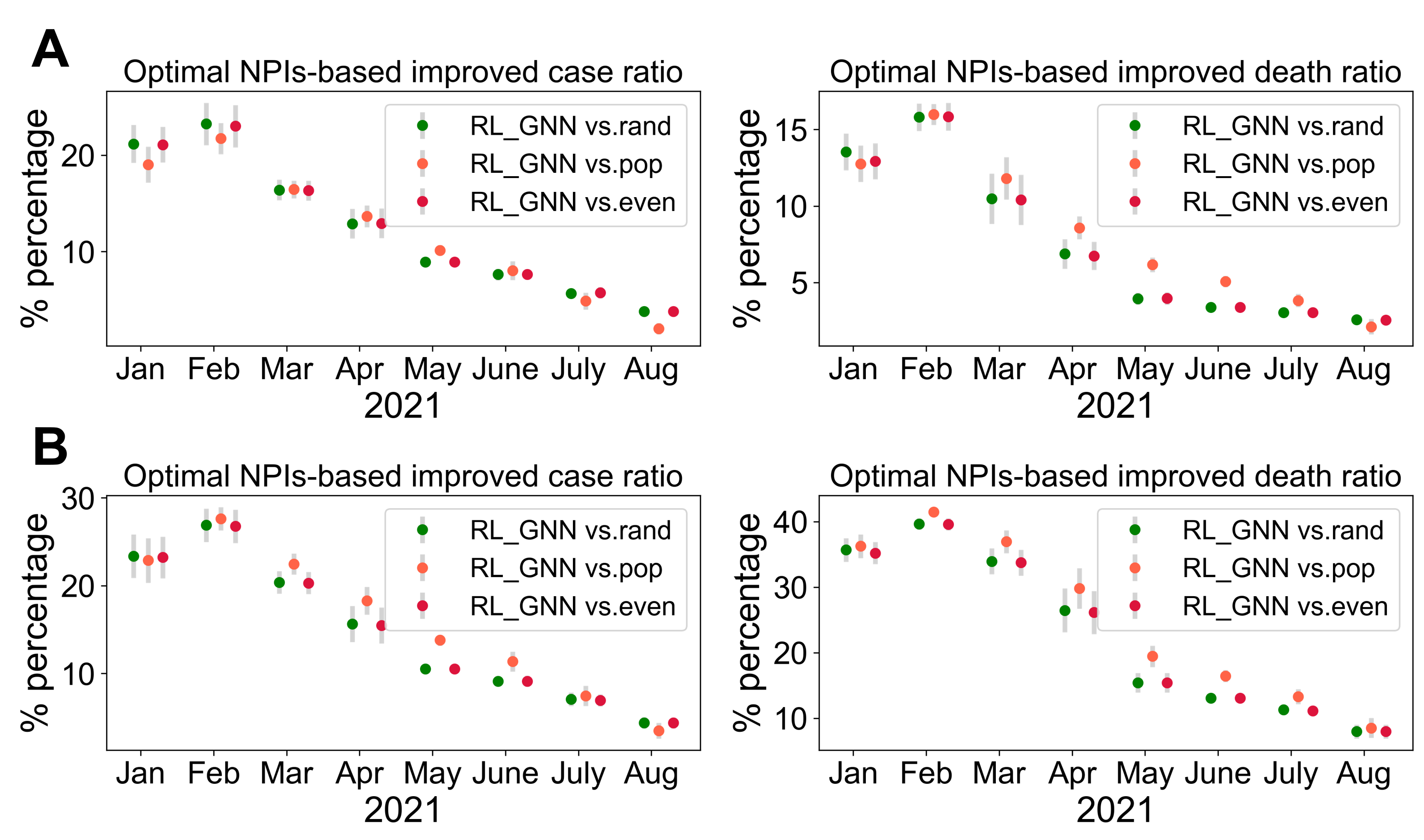}
    \caption{The time series evaluation of RL-GNN framework under the age-based mobility restriction --$IRD_{pro}$ and $D_{pro}$; A: restricting cross-zone mobility for the top 10\% oldest zones; B: restricting cross-zone mobility for the top 10\% youngest zones.}
    \label{fig:RL_GNN_age_timeseries} \vspace{-2mm}
\end{figure}

\begin{figure}[t]
    \centering
    \includegraphics[width=\linewidth]{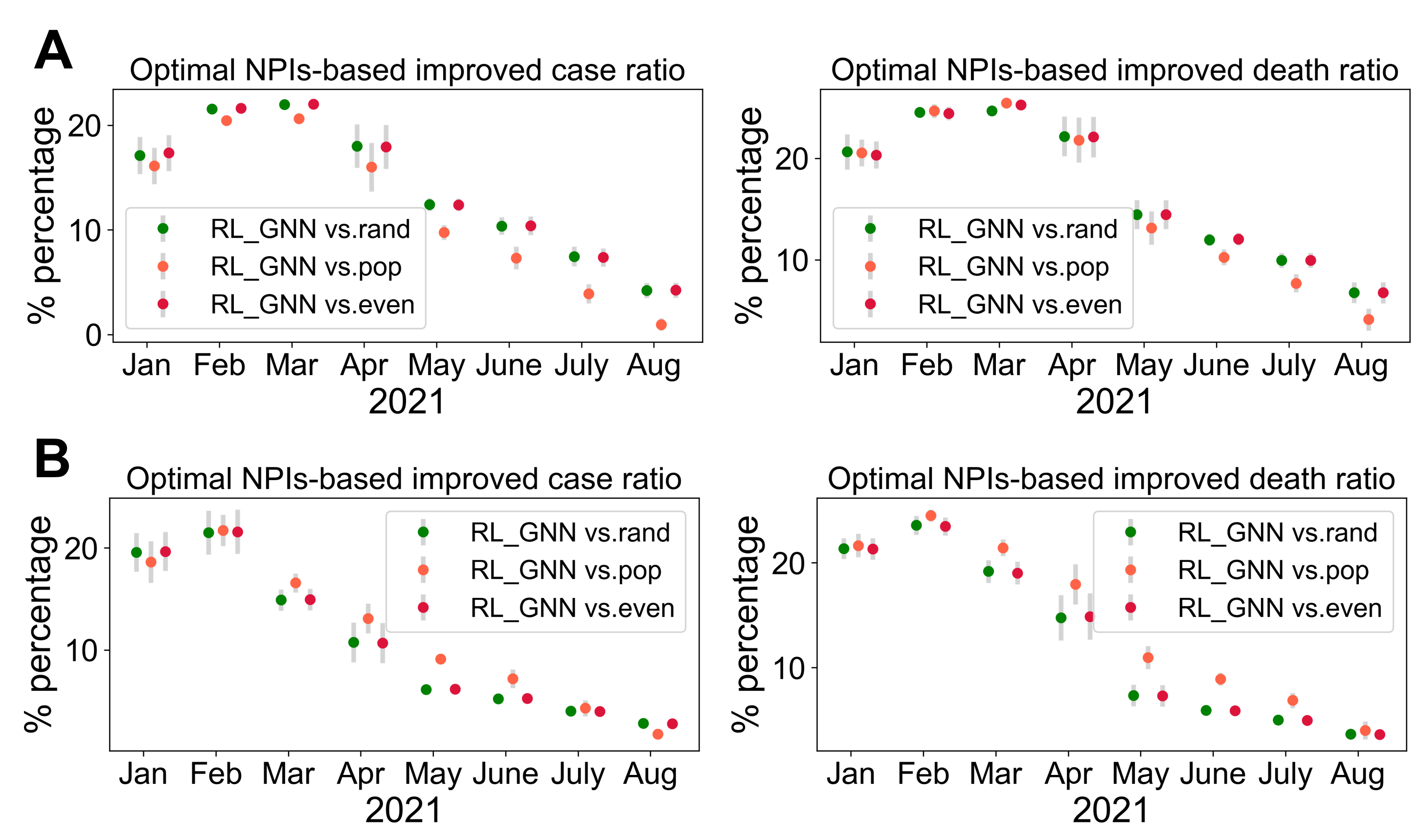}
    \caption{The time series evaluation of RL-GNN framework under the transit usage restriction --$IRD_{pro}$ and $D_{pro}$; A: low and median transit usage restriction; B: median capacity transit usage restriction.}
    \label{fig:RL_GNN_transit_timeseries} \vspace{-2mm}
\end{figure}

In addition, Figures~\ref{fig:RL_GNN_income_timeseries}, \ref{fig:RL_GNN_age_timeseries}, and \ref{fig:RL_GNN_transit_timeseries} visualize the optimal NPIs-based improved case and death ratios for each NPI from January to August 2021. We observe that the optimal NPIs-based improved case and death ratios significantly vary along with time. Therefore, we reject the fourth hypothesis.


 The study has several limitations: $(i)$ the proposed Trans-vaccine-SEIR model is not appropriate to predict long-term disease dynamics since external factors, such as NPIs, would significantly change the disease propagation state. We address this issue by dividing the study period into eight periods when calibrating the disease parameters and simulating the eight periods continuously. $(ii)$ due to the lack of reported population in separating $S$ and $E$ compartments, the initial $S,E,I,R,D$ population are based on the estimation from the literature. $(iii)$ the mobility patterns in diverse NPIs scenarios are approximated by mobile phone data, which might deviate from the realistic scenario. However, it would not affect the quality of robustness examination of the RL-GNN framework since the optimal vaccine allocation strategy could be adaptive to the input mobility pattern and adjusted as environment changes.

\section{Conclusion}

This paper proposes a framework for effective vaccine prioritization at the micro-geographical level to reduce the overall burden of the pandemic when the vaccine supply is limited. 
Specifically, we propose a Trans-vaccine-SEIR model that improves the complex disease propagation simulation from the epidemiological and non-epidemiological aspects by integrating the effects of the census block-level mobility dynamics. 
Using the Trans-vaccine-SEIR model as the environment simulator, we present an RL-GNN framework to learn an optimal vaccine allocation strategy in a high-degree spatiotemporal disease evolution environment instead of achieving a sub-optimal vaccine allocation strategy under simplified assumptions. 
The evaluation examines the simulator accuracy and shows that the optimal vaccine allocation strategy from the RL-GNN framework is significantly more effective than the baseline strategies. The extensive evaluations based on multiple NPIs verify the robustness of the proposed framework with diverse mobility patterns. In particular, we find the optimal vaccine allocation strategy in the transit usage restriction scenario is more effective in reducing infections and deaths than cross-zone mobility restrictions for the top 10\% age-based and income-based zones.
\bibliographystyle{ACM-Reference-Format}
\bibliography{sample-base}

\appendix

\section{GNN-based agent module} \label{sec.GNN-agent-module}

\begin{figure}[t]
\centering
\includegraphics[width=\linewidth]{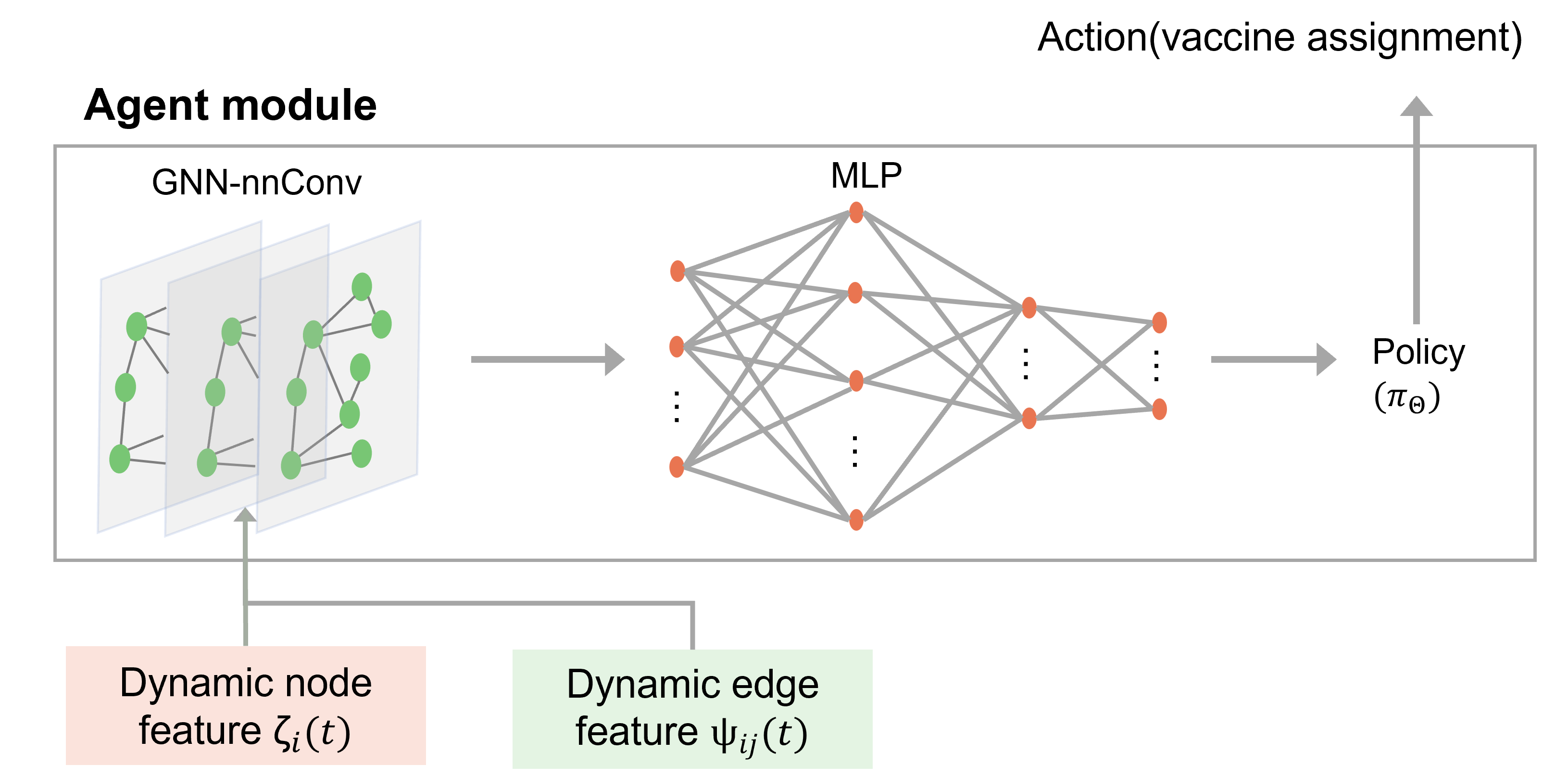}
\caption{GNN-based agent module. We apply GNN to extract features from the environment graph. Based on the features, we then apply MLP to predict policy for vaccine assignment. }
\label{fig:agent_module}
\end{figure}

GNNs\cite{zhou2020graph} are neural models that capture the dependence of graphs via message passing between the nodes of graphs. It has the ability to extract multi-scale spatial features and compose them to construct highly expressive representations. As mobility based disease evolution in the census block level can be transferred to the node and edge (mobility connection) in the evolving graph. Instead of the multilayer perceptron (MLP), GNN has its superiority to capture the variants of each components. 
Within the agent module, we use the GNN to extract the node and edge features in the environment setting (graph) in the Trans-vaccine-SEIR module. 
The GNN-based agent module presents in Figure \ref{fig:agent_module}. We use a PPO optimizer, an actor-critic algorithm in the RL framework. It requires a critic module to estimate the value function in a given state. We use the GNN-nnConv to extract the feature from the observation states. Then, we construct the critic using MLP to the agent module. The dynamic node feature $\zeta_i (t)$ including $S^u_i,S^v_i,E^u_i,E^v_i,I^u_{ai},I^u_{si},I^v_{ai},I^v_{si},R^u_i,R^v_i,V^u_i$. 
and the each edge features  $\Psi_{ij}(t)$ including $S^u_{ij},S^v_{ij},E^u_{ij},E^v_{ij},I^u_{a,ij},I^v_{a,ij},I^u_{s,ij},I^v_{s,ij},R^u_{ij},R^v_{ij}$. 
We apply a softmax operation on the MLP output of the agent module to output the importance distribution for vaccine allocation. 

The graph's edge (mobility connection between zones) plays an essential role in disease propagation. 
To extract the edge features effectively, the convolution layer for our GNN is nnConv~\cite{gilmer2017neural}. 
That is, in the message passing phase, hidden states $h^t_v$ at each node in the graph are updated based on messages $m^{t+1}_v$ according to the function:

\begin{equation}
\begin{split}
    m^{t+1}_i = \mathcal{N}(i)M_t(h^t_i,h^t_j,e_{ij}) = A_{e_{ij}}h_j \\
    h^{t+1}_v = U_t(h^t_v,m^{t+1}_v)
\end{split}
\end{equation}

Where $M_t$  is a message function and $U_t$ is the vertex update function. $A(e_{vw})$ is a neural network which maps the edge vector $e_{ij}$ to a $d\times d$ matrix. 
In practice, $A(e_{vw})$ is a MLP. 
We can control the GNN's expressive power by changing the model size of the MLP accordingly. 

\section{Trans-vaccine-SEIR model calibration} \label{sec.Trans-vaccine-SEIR-model-calibration}
\begin{figure}[t]
\centering
\includegraphics[width=\linewidth]{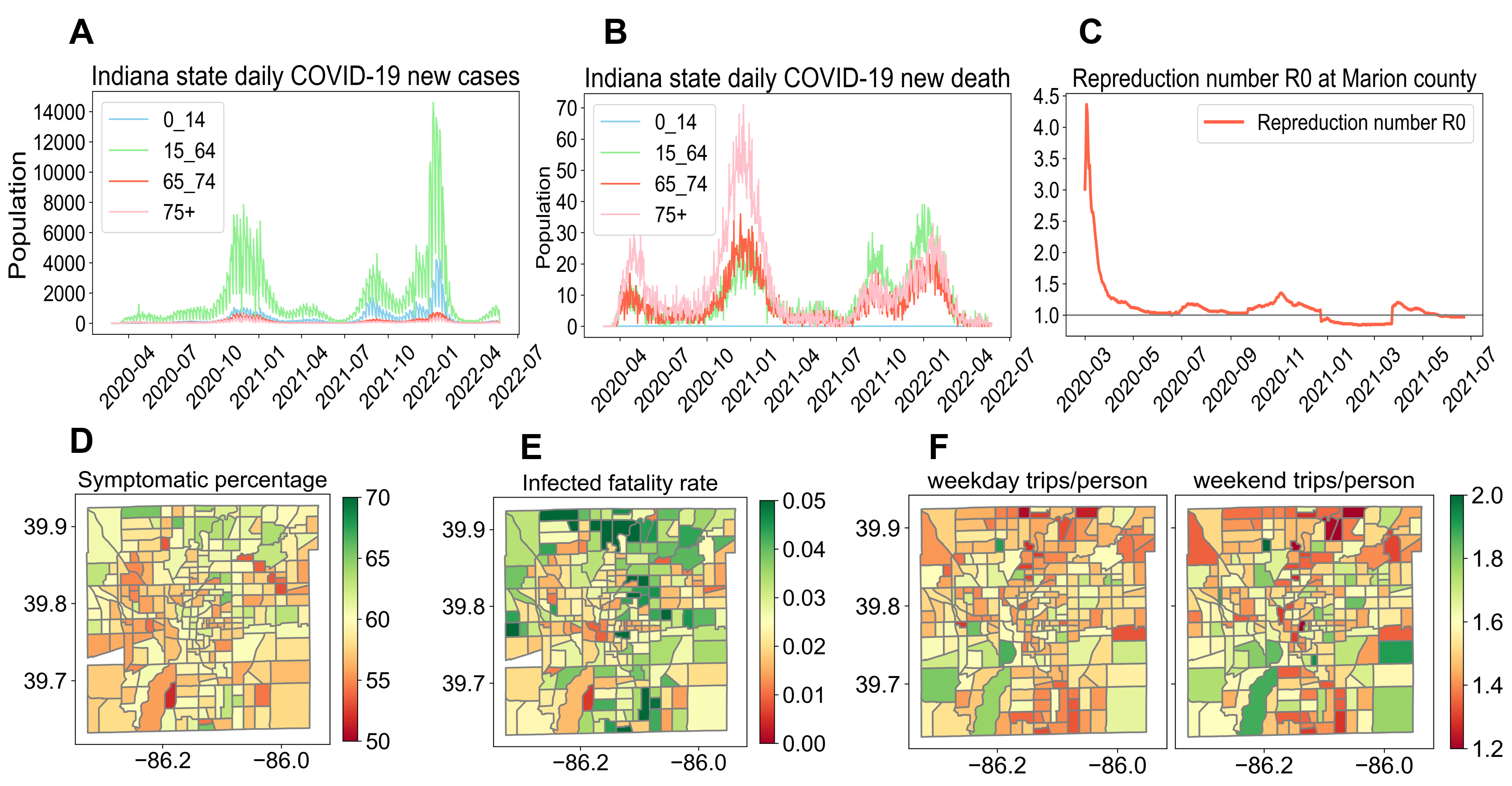}
\caption{Parameter visualization. A is the daily new cases in Indiana state. B is the daily new deaths in Indiana state. C is the reproduction number $R_0$ at Marion county. D is the percentage of infected individuals presenting symptoms. E is the IFR in Marion county. F is the spatial variation of trips per person. }
\label{fig:DISEASE_PARA}
\end{figure}

Parameter visualization can be seen in Figure\ref{fig:DISEASE_PARA}.
The summary of the parameter description and source can be found in Table~\ref{tab:parameter_source}.

\begin{table}[t]
    \centering
    \begin{tabular}{{p{0.08\textwidth}p{0.2\textwidth}p{0.13\textwidth}}}
    \hline
         Notation& Value & Source\\ \hline
         $N_i$ & US census block  & \cite{UScensus}  \\
         $S_{i}^t$ & calibrated. The same for $E$, $I$, and $R$ population & \cite{timmurphy1} \\
         $D$ & 2 & \cite{indianacensusreport} \\
         $\beta^a$ & 0.016 - 0.03 & fitted \\
         $\beta^t$ & $\beta_{d=1}^t \in (0.006-0.002),\beta_{d=2}^t \in (0.008-0.02)$ &fitted  \\
         $c^d$ & $c_{d=1} = 0.8925, c_{d=2} = 0.1075 $ & \cite{indianacensusreport} \\
          $\frac{1}{\sigma}$ & $1/3 -1/6$  & \cite{anderson2020reproduction}\\
          $\frac{1}{\gamma_e}$ & $\frac{1}{\gamma_s} \in (22,30)$ days and $\frac{1}{\gamma_a} \in (5,14)$ days & \cite{anderson2020reproduction}\\
          $k_{ij}$ & calibrated & mobile phone data \\
          $k_{ij,kl}^d$ & calibrated & mobile phone data \\
          $g_i,m_{ij}, r_{ij}$ & calibrated &  mobile phone data\\
          $\delta$ & 90\% & \cite{thompson2021interim}\\
          $q$ & see Figure \ref{fig:DISEASE_PARA} E & fitted \\
          $IFR$ & see Figure \ref{fig:DISEASE_PARA} D & fitted \\
          $f_i, f_{ij}$ & calibrated by equations \ref{equ:activity1} and \ref{equ:activity2} & fitted\\
          $h_{\overleftarrow{ij}}, h_{\overrightarrow{ij}}$ & calibrated by equations \ref{equ:travel1} and \ref{equ:travel2} & fitted \\
          $f_a^u$ & 0-1& \cite{cdcandaspr}\\
          $f_s^v$ & 0.5 & \cite{islam2021evaluation}\\ \hline
    \end{tabular}
    \caption{Model parameters and sources}
    \label{tab:parameter_source}
\end{table}

\section{Environment Simulation} \label{sec.pseudo-code}
The pseudo-code for the Trans-vaccine-SEIR model in the environment is shown in Algorithm \ref{alg:cap}. 

\begin{algorithm}[H]
\caption{Environment space}\label{alg:cap}
\begin{algorithmic}
\Require $g_i, m_{ij}$ \;\\
$g_i$ is the departure rate \;\\
$m_{ij}$ is the movement rate between $i$ and $j$ \;
\Require $m,n \geq 0$ \; \\
$m$ is the amount of vaccine supply\\
$n$ is $S^u$, the non-vaccinated susceptible population \;
\Require $n \geq m$ \; 
\State$N\gets m$\;\Comment*[r]{Initialization}
\State$s_i,e_i,i_i,r_i,d_i \gets s^0_i,e^0_i,i^0_i,r^0_i,d^0_i$ 
\While{$N \leq n$}    \Comment*[r]{Beginning of day}
\State $s_i \gets s_i+\delta s_i$\; \Comment*[r]{Action}
\State $s_{ij},e_{ij},i_{ij},r_{ij} \gets g_i,m_{ij},s_i,e_i,i_i,r_i$\; \Comment*[r]{Mobility}
\State $f_i \gets \delta f_i$\; 
\State $f_{ij} \gets \delta f_{ij}$\; 
\State $h_{\overleftarrow{ij}} \gets \delta  h_{\overleftarrow{ij}}$\;
\State $h_{\overrightarrow{ij}} \gets \delta h_{\overrightarrow{ij}}$\;  
\State$s_i \gets s_i+\Delta s_i$\; \Comment*[r]{Update state}
\State$e_i \gets e_i+\Delta e_i$\;
\State$i_i \gets i_i+\Delta i_i$\;
\State$r_i \gets r_i+\Delta r_i$\;
\State$d_i \gets d_i+\Delta d_i$\; \Comment*[r]{End of day}

\EndWhile \\
\Return $s_i,e_i,i_i,r_i,d_i,s_{ij},e_{ij},i_{ij},r_{ij},d_{ij}$\; 
\end{algorithmic}
\end{algorithm}

\end{document}